\documentclass[twocolumn]{openjournal}

\usepackage{lipsum}

\usepackage{xcolor}
\usepackage{textgreek}
\usepackage[utf8]{inputenc}
\usepackage[english]{babel}

\usepackage{hyperref}
\hypersetup{
    unicode, 
    colorlinks=true,
    linkcolor=linkcolor,
    citecolor=linkcolor,
    filecolor=linkcolor,
    urlcolor=linkcolor,
}
\usepackage{color,colortbl}
\definecolor{linkcolor}{rgb}{0.0,0.3,0.5}
\usepackage{tensind}
\tensordelimiter{?}
\DeclareGraphicsExtensions{.bmp,.png,.jpg,.pdf}
\usepackage{verbatim}
\usepackage[normalem]{ulem}
\usepackage{orcidlink}
\usepackage{soul}

\graphicspath{ {./figs/} }

\begin{document}
\title{Collision between molecular clouds IV: The role of feedback and magnetic field in head on collisions}

\author{Tabassum S. Tanvir \orcidlink{0000-0002-0862-0701}}

\email{ttanvir@iastate.edu}
\affiliation{Department of Physics and Astronomy, Iowa State University, 2323 Osborn Rd, Ames, IA 50011, USA}
\affiliation{Research School of Astronomy and Astrophysics, Australian National University, Canberra, ACT 2611, Australia}

\author{Michael Y. Grudi\'{c} \orcidlink{0000-0002-1655-5604}}
\affiliation{Center for Computational Astrophysics, Flatiron Institute, 162 5th Ave, New York, NY 10010, USA}


\begin{abstract}
We systematically investigate how cloud–cloud collisions influence star formation, emphasizing the roles of collision velocity, magnetic field orientation, and radiative feedback. Using the first cloud-cloud collision simulations that model individual star formation and accretion with all stellar feedback mechanisms, we explore the morphological evolution, star formation efficiency (SFE), fragmentation, stellar mass distribution, and feedback-driven gas dispersal. Our results show that cloud collisions substantially enhance the rate and timing of star formation compared to isolated scenarios, though the final SFE remains broadly similar across all setups. Lower collision velocities facilitate the prolonged gravitational interaction and accumulation of gas, promoting sustained star formation characterized by elongated filamentary structures. Conversely, high-velocity collisions induce rapid gas compression and turbulent motions, leading to intense but transient episodes of star formation, which are curtailed by feedback-driven dispersal. The orientation of the magnetic field markedly affects collision outcomes. Parallel fields allow gas to collapse efficiently along magnetic lines, forming fewer but more massive stars. In contrast, perpendicular fields generate significant magnetic pressure, which stabilizes the shock-compressed gas and delays gravitational collapse, resulting in more distributed and less massive stellar fragments. Radiative feedback from massive stars consistently regulates star formation, halting further gas accretion at moderate efficiencies (10–15\%) and initiating feedback-driven dispersal. Although the cloud dynamics vary significantly, the stellar mass function remains robust across scenarios—shaped modestly by magnetic orientation but only weakly influenced by collision velocity.
\end{abstract}


\maketitle

\section{Introduction}
Giant Molecular Clouds (GMCs) are the primary sites of star formation within galaxies, where stars form across the entire mass spectrum, ranging from low-mass stars to massive stellar clusters. However, the precise mechanisms governing how GMCs convert gas into stars, particularly massive stars, remain topics of active investigation. One significant proposed mechanism for enhancing star formation efficiency and triggering massive star formation is collisions between GMCs \citep{2009ApJ...696L.115F, 2014ApJ...780...36F, 2015ApJ...806....7T, 2015MNRAS.446.3608D,2020MNRAS.494..246T,2021MNRAS.506..824T}. Such cloud-cloud collisions (CCCs) provide unique environments characterized by highly compressed, dense gas layers capable of rapidly initiating gravitational collapse, potentially forming massive stars and stellar clusters \citep{1992PASJ...44..203H,2014ApJ...792...63T,2015MNRAS.453.2471B,2023MNRAS.522.4972S}.

Observationally, many massive star-forming regions within the Milky Way and nearby galaxies exhibit kinematic signatures indicative of recent cloud collisions, including broad velocity bridges, overlapping molecular components with distinct velocities, and high-density shocked layers. Prominent examples include the Westerlund 2 cluster \citep{2009ApJ...696L.115F, 2010ApJ...709..975O}, NGC 3603 \citep{2014ApJ...780...36F}, and RCW 120 \citep{2015ApJ...806....7T}, all of which exhibit massive star clusters coincident with collision sites. Additional compelling cases are observed in regions such as the Trifid Nebula (M20; \citealt{2011ApJ...738...46T}), RCW 38 (\citealt{2016ApJ...820...26F}), and M17 (\citealt{2018PASJ...70S..42N}), each displaying clear velocity offsets indicative of colliding molecular clouds. The starburst cluster NGC 6334, studied extensively by \citet{2018PASJ...70S..41F}, also shows molecular gas kinematics consistent with energetic cloud collisions. More recently, the massive cluster-forming regions W49A \citep{2022PASJ...74..128M} and G333.6-0.2 \citep{2025AJ....169..181K} were identified as likely products of high-velocity cloud interactions, reinforcing the association between massive star formation and cloud-cloud collision phenomena. Together, these observational results strongly support the hypothesis that cloud-cloud collisions constitute a pivotal mechanism, potentially accelerating and enhancing massive star formation across diverse galactic environments.

From a theoretical standpoint, numerical simulations have extensively examined the role of cloud collisions in facilitating star formation, investigating various collision parameters including relative velocity, impact parameter, cloud density profiles, and initial gravitational boundedness \citep{2014ApJ...792...63T, 2015MNRAS.453.2471B,2017MNRAS.465.3483B, 2017ApJ...835..137W, 2017ApJ...841...88W, 2020MNRAS.499.1099L, 2020MNRAS.494..246T,2021MNRAS.506..824T}. Early hydrodynamic studies established that collisions significantly enhance gas density in shocked interfaces, forming filamentary structures and dense cores conducive to star formation. \citet{2020MNRAS.494..246T, 2021MNRAS.506..824T} particularly emphasized the importance of cloud virial states, demonstrating that gravitationally bound clouds produce more pronounced filamentary networks and higher star formation efficiencies compared to unbound collisions.

Despite substantial progress, the inclusion of magnetic fields and detailed stellar feedback physics has remained relatively limited, leaving significant uncertainties regarding their roles in cloud collisions. Magnetic fields, ubiquitous within GMCs, profoundly influence cloud dynamics by providing additional pressure support, guiding gas flows, and potentially suppressing or promoting fragmentation depending on their orientation and strength \citep{2013ApJ...774L..31I, 2017ApJ...841...88W, 10.1093/mnras/stab150}. Magnetohydrodynamic (MHD) simulations indicate that magnetic fields aligned parallel to collision axes allow efficient compression and rapid gravitational collapse, while perpendicular fields can stabilize post-shock gas layers against immediate fragmentation \citep{2017ApJ...841...88W, 2023MNRAS.522.4972S}. Moreover, stellar feedback—particularly radiative feedback from massive stars—critically regulates the star formation process by heating, ionizing, and dispersing the surrounding gas, thereby limiting star formation efficiencies \citep{2018MNRAS.481.2548G, 2018ApJ...859...68K, 2021MNRAS.506.2199G}. Prior studies incorporating stellar feedback in collision scenarios found substantial moderation of star formation compared to purely hydrodynamic models \citep{2020MNRAS.497.3830F}. However, these studies typically used simplified treatments, neglecting detailed radiative transfer or magnetic fields, highlighting the need for comprehensive simulations. To date, no study has simultaneously modelled cloud–cloud collisions with the full combination of magnetohydrodynamics, multi-frequency radiative transfer, and protostellar evolution required to follow star formation self-consistently down to individual stars. Existing simulations either lack the resolution to capture fragmentation accurately or omit key feedback channels—limitations that prevent a complete assessment of how collision velocity, magnetic geometry, and stellar radiation jointly shape star formation outcomes. The STARFORGE framework \citep{2021MNRAS.506.2199G} overcomes these constraints by coupling ideal MHD with detailed radiative physics at sub-solar-mass resolution, enabling us to track the interplay between shock compression, magnetic support, and radiative heating in a regime previously inaccessible.

In this paper, we present high-resolution ideal magnetohydrodynamic simulations of head-on collisions between turbulent molecular clouds, systematically varying collision speed, magnetic field orientation, and explicitly including radiative stellar feedback. Utilizing the STARFORGE framework \citep{2021MNRAS.506.2199G}, which couples detailed radiative transfer with magnetohydrodynamics, we explore how these critical parameters collectively influence cloud morphology, star formation efficiency, stellar mass distribution, and gas dispersal dynamics.

\section{Numerical method and Initial condition}
\subsection{Numerical method}

Our simulations use the STARFORGE framework \citep{2021MNRAS.506.2199G}, which couples ideal magnetohydrodynamics (MHD), protostellar evolution, and multi-channel stellar feedback within the GIZMO code. Gas dynamics are evolved with the Meshless Finite Mass (MFM) method \citep{2015MNRAS.450...53H}, a mesh-free, Lagrangian finite-volume scheme that provides excellent conservation properties and Galilean invariance while accurately capturing highly supersonic, turbulent flows. Magnetic fields are evolved using GIZMO’s constrained-gradient MHD solver \citep{2016MNRAS.455...51H}, ensuring stable and divergence-controlled field evolution during gravitational collapse.

Thermodynamics follow the FIRE-3 ISM heating, cooling, and chemistry model \citep{2023MNRAS.519.3154H}, which includes metal-line cooling, molecular and atomic cooling, photoelectric heating, cosmic-ray–motivated heating prescriptions, and a redshift-dependent UV background. This treatment yields realistic thermal behaviour across the full density and temperature ranges encountered in molecular clouds.

Stars are represented by sink particles that form when gas exceeds the Jeans resolution threshold and satisfies local gravitational binding and collapse criteria. While the method does not prevent artificial fragmentation strictly, the STARFORGE sink formation algorithm ensures that unresolved collapse scales are treated at the sub-grid level, producing robust stellar mass distributions \citep{2021MNRAS.506.2199G}. Once formed, protostars evolve following the STARFORGE protostellar model, which self-consistently tracks accretion, luminosity evolution, and radius contraction.

Radiative processes are handled self-consistently through GIZMO's radiative transfer solver, tracking multiple frequency bands including infrared, optical, near-UV, far-UV, and ionizing radiation bands. Protostellar jets and stellar winds are included following \citet{2021MNRAS.506.2199G}, injecting momentum and energy into the surrounding gas and playing a key role in regulating fragmentation and driving turbulent dispersal. These mechanical and radiative feedback processes act in concert with magnetic fields to shape the evolution of the shocked layer and regulate star formation.

Magnetic fields are initialized uniformly in each cloud with a mass-to-flux ratio of $\mu = 1.3$. We explore two field orientations: parallel, where magnetic fields align with the collision axis and allow efficient channeling of gas into the shock, and perpendicular, where fields oppose compression and increase magnetic pressure support in the post-shock layer.

\subsection{Initial condition}
\label{sec:Num} 

Each simulation initializes two spherical molecular clouds, each with a mass of $M_0 = 2000\,M_{\odot}$ and a radius of $R_{\rm cloud} = 3$\,pc. The cloud has uniform density and is seeded with a random Gaussian turbulent velocity field, characterized by a power spectrum $E(k) \propto k^{-2}$. This sets the turbulent virial parameter, which is defined by the following equation
\begin{equation}
\alpha_{\rm turb} \equiv \frac{5 \| \mathbf{v}_{\rm turb} \|^2 R_{\rm cloud}}{3 G M_0} =2,
\label{eqnvir}
\end{equation}
where $R_{\rm cloud}$ and $\rm M_{0}$ are the radius and mass of the cloud. A uniform magnetic field threads each cloud with a strength set by a mass-to-flux ratio of $\mu = 1.3$, corresponding to a slightly supercritical configuration. We explore two configurations:

(i) In the parallel case, the magnetic field is aligned with the collision axis, such that the bulk motion of the clouds occurs along the field lines;

(ii) In the perpendicular case, the magnetic field is orthogonal to the collision axis, so that the clouds move across the field lines during the encounter.

These initial conditions are generated using the MAKECLOUD tool, as described in \citet{2022MNRAS.515.4929G}. The cloud is placed in a large computational domain, surrounded by a tenuous ambient medium with a density contrast of 1:1000. No external forcing is applied after setup, allowing the initial turbulence to decay naturally. The freefall is fixed across all the simulations given the clouds are identical, for a cloud with mass 2000 $\rm M_{\odot}$ and radius of 3 pc the freefall time is,
\begin{equation}
t_{\mathrm{ff}} = \sqrt{\frac{3\pi}{32 G \rho_{0}}} = 1.3 Myr 
\label{eqnff}
\end{equation}
where $\rho_{0}$ is the density of the cloud. The cloud is discretized using  $2 \times 10^6$ cells per cloud, corresponding to a minimum cell mass of $\sim 10^{-3}\,M_{\odot}$. This resolution ensures that local Jeans lengths are resolved, preventing artificial fragmentation. Gas thermodynamics and radiation are treated self-consistently following the STARFORGE methodology, employing cooling and heating rather than an isothermal assumption. The gas is initially cold ($T = 20$\,K) but equilibrates with a uniform interstellar radiation field, accounting for gas--dust thermal coupling under solar-neighborhood conditions. Unlike \citet{2020MNRAS.494..246T}, who used a fixed 30\,K barotropic equation of state, our simulations allow gas temperature to evolve dynamically in response to compressional heating, cooling, and radiative feedback. No star particles are present initially, but radiative transfer ensures that protostellar heating and radiation from forming stars are accurately modelled.

Collisions are formed between two identical clouds (mass ratio: 1), following the framework of \citet{2020MNRAS.494..246T}. The clouds are placed on a collision course with a centre-to-centre separation of $3R_{\rm cloud} \approx 9$\,pc and assigned equal and opposite bulk velocities along a defined collision axis. We explore two relative collision speeds: a slow case with $v_{\rm coll} = 1\,\sigma_{\rm turb}$ and a fast case with $v_{\rm coll} = 5\,\sigma_{\rm turb}$, where $\sigma_{\rm turb} \approx 3$\,km\,s$^{-1}$ is the 3D rms turbulent velocity. These correspond to relative speeds of approximately 3\,km\,s$^{-1}$ and 15\,km\,s$^{-1}$, respectively. The collision time, $t_{\rm coll}$, defined as the moment when the cloud surfaces make contact $t_{c}$ is defined by , 

\begin{equation}
t_{\mathrm{c}} = \frac{D -2R}{v_{coll}} = 0.9, 0.2 Myr 
\label{eqnff}
\end{equation}
where \( D \) is the initial centre-to-centre separation between the clouds, \( R \) is the cloud radius, and \( v_{\mathrm{coll}} \) is the relative collision velocity. For clouds with \( D = 9\,\mathrm{pc} \) and \( R = 3\,\mathrm{pc} \), this yields \( t_{\mathrm{c}} \approx 0.9\,\mathrm{Myr} \) for the slow collision case (\( v_{\mathrm{coll}} = 3\,\mathrm{km\,s^{-1}} \)) and \( t_{\mathrm{c}} \approx 0.2\,\mathrm{Myr} \) for the fast collision case (\( v_{\mathrm{coll}} = 15\,\mathrm{km\,s^{-1}} \)). The collision timescales are compared to the clouds' initial turbulent crossing time and free-fall time. In fast collision runs, $t_{\rm coll}$ is short relative to the free-fall time ($t_{\rm coll} \ll t_{\rm ff}$), limiting pre-collision gravitational collapse. In contrast, slow collisions occur on timescales comparable to or longer than the free-fall time ($t_{\rm coll} \gtrsim t_{\rm ff}$), allowing denser regions to begin collapsing and potentially forming stars before impact. This distinction between prompt and delayed collisions, as analysed by \citet{2020MNRAS.494..246T}, significantly influences the collision outcome. \autoref{tab:ic_summary} summarises the key initial parameters for all simulations.

\begin{table*}
    
\centering
\caption{Summary of initial conditions for all simulations. Each cloud has mass $M = 2000\,M_\odot$, radius $R = 3\,\mathrm{pc}$, turbulent virial parameter $\alpha = 2$, and mass-to-flux ratio $\mu = 1.3$. The clouds are initially separated by $D = 3R$ and collide head-on ($b = 0$). The table lists runs for two different magnetic field orientations (parallel and perpendicular to the collision axis) and two different collision speeds in units of the internal turbulent velocity $v_{\rm turb}$.}
\label{tab:ic_summary}
\begin{tabular}{ccccccccccc}
\hline
Run & $M\, (M_\odot)$ & $N_{\rm cell}$ & $\Delta m\, (M_\odot)$ & $R$ (pc) & $\alpha$ & $\mu$ & $D/R$ & $b/R$ & $v_{\rm coll}/v_{\rm turb}$ & Field \\
\hline
1x perpendicular & 2000 & $2\times10^6$ & $10^{-3}$ & 3 & 2 & 1.3 & 3 & 0 & 1 & $\perp$ \\
5x perpendicular & 2000 & $2\times10^6$ & $10^{-3}$ & 3 & 2 & 1.3 & 3 & 0 & 5 & $\perp$ \\
1z parallel      & 2000 & $2\times10^6$ & $10^{-3}$ & 3 & 2 & 1.3 & 3 & 0 & 1 & $\parallel$ \\
5z parallel      & 2000 & $2\times10^6$ & $10^{-3}$ & 3 & 2 & 1.3 & 3 & 0 & 5 & $\parallel$ \\
\hline
\end{tabular}
\end{table*}

\section{Results}
This section is divided into three subsections. The first subsection discusses the morphology of the collision product; the next subsection examines the star formation rate and efficiencies. In the third subsection, we discuss whether the clouds are bound or not as a result of the collision. 

\subsection{Morphology of the clouds}

\begin{figure*}
    \centering
    \includegraphics[width=2.0\columnwidth]{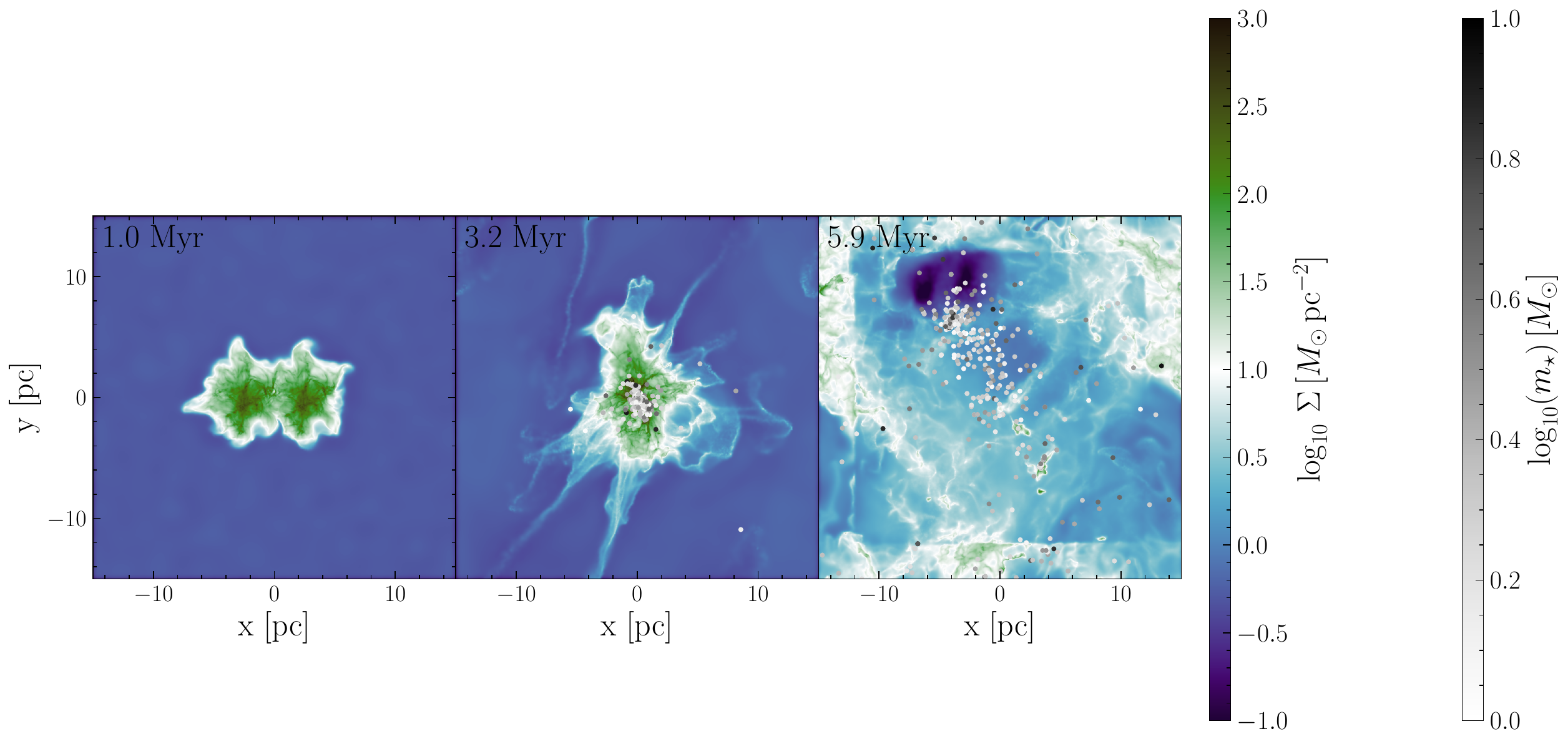}
    \caption{Projected gas surface density evolution for the 1x perpendicular magnetic field collision scenario at three representative evolutionary stages (early, intermediate, and late times). Each panel highlights how the perpendicular magnetic field shapes the morphology, initially stabilizing the gas layer against fragmentation and ultimately facilitating a dramatic feedback-driven dispersal in later stages.}
    \label{fig:1xproj}
\end{figure*}

\textbf{1x perpendicular}:

In this simulation, the collision speed is 3 $\mathrm{km\,s^{-1}}$, and the magnetic field is perpendicular to the collisional axis. From \autoref{fig:1xproj} we see that at the early stage the two clouds make initial contact, forming a compressed, irregular gas layer at the interface with dense filaments and clumps where the first stars begin to form; star particles are sparse and embedded in the densest concentrations. By the intermediate stage, the collision region has evolved into a complex network of dense filaments and compact cores, with a protocluster forming in the central area. The gas remains a contiguous shocked slab, and star particles exhibit moderate clustering within the brightest regions, with small, localized, low-density pockets appearing around young stars. At the late stage, stellar feedback has reshaped the morphology into a large low-density cavity around a highly clustered central star group, transforming the gas from a dense, continuous slab into a hollowed-out structure with remnant filaments and shells swept to the cavity’s edges; bright stars sit at the cavity’s center with additional stars along the rim, marking a cavity-dominated configuration in stark contrast to the earlier filamentary phase.

\begin{figure*}
    \centering
    \includegraphics[width=2.0\columnwidth]{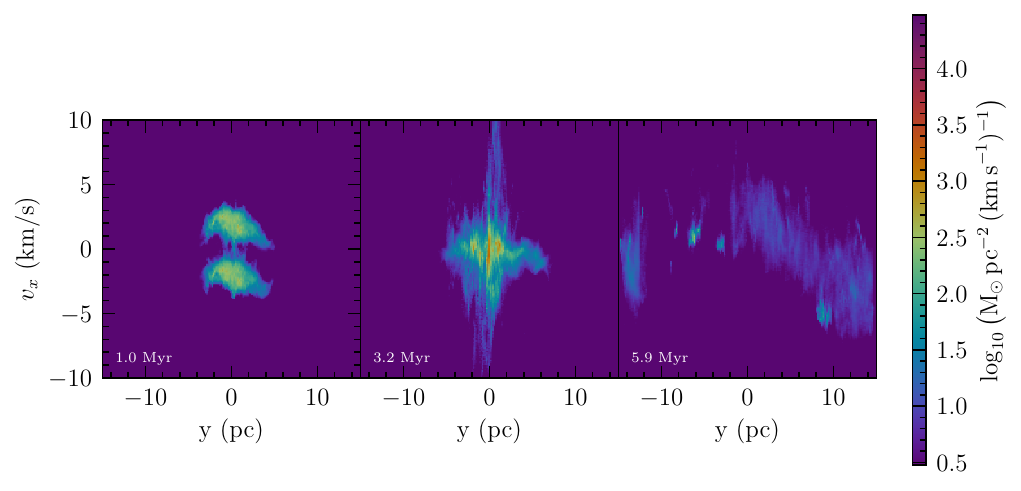}
    \caption{Position–velocity diagrams for the 1x perpendicular magnetic field collision scenario at the same evolutionary stages depicted in \autoref{fig:1xproj}. These diagrams illustrate how the velocity structure evolves from two initially distinct clouds through turbulent mixing in the intermediate stage to a highly dispersed, feedback-influenced velocity field at late times, highlighting the complex interplay between magnetic fields, gas dynamics, and stellar feedback.}
    \label{fig:1xpv}
\end{figure*}
Since the surface density plots don’t contain much dynamical information, we now turn towards the position-velocity plots for this simulation. From \autoref{fig:1xpv} we can see that at an early stage, the two molecular clouds make initial contact, forming a compressed, irregular gas layer with dense filaments and clumps where the first stars begin to emerge. The PV plot shows two distinct, symmetric high-density clumps centered around \(y\approx0 \) pc, with velocities from -5 to +5 $\mathrm{km\,s^{-1}}$ connected by a bridge feature between them very similar to bridge features seen in \citep{2015MNRAS.454.1634H}. At the intermediate stage, the collision region has become a complex network of dense filaments and compact cores, with a protocluster in the central area; the gas remains a contiguous shocked slab, and star particles show moderate clustering within the brightest regions, with small, localized low-density pockets appearing around young stars. At the late stage, stellar feedback has reshaped the morphology into a large low-density cavity around a highly clustered central star group, transforming the gas from a dense, continuous slab into a hollowed-out structure with remnant filaments and shells swept to the cavity’s edges; bright stars sit at the cavity’s center with additional stars along the rim, marking a cavity-dominated configuration in stark contrast to the earlier filamentary phase.

\textbf{5x perpendicular}:

In this simulation, the collision speed is 15 $\mathrm{km\,s^{-1}}$ and the magnetic field is perpendicular to the collisional axis. At the early stage, the two clouds collide violently and produce a thin, sharply bounded midplane interface: a turbulent post-shock slab with only minor fragmentation. The transverse field is strongly compressed, injecting magnetic pressure and tension that resist lateral compression, channel motions within the layer, and suppress immediate collapse—no star formation is yet visible. By the intermediate stage, the compressed layer has become a disordered, filament-rich sheet with complex substructure. Star formation has begun, but sink particles remain sparsely distributed and weakly clustered; the perpendicular field continues to regulate flows, favoring tangled filaments and fragmented clumps over a coherent, dense core. Stellar feedback is present but modest, and the slab stays contiguous without a dominant central condensation. At the late stage, the morphology changes little: the post-shock layer remains filamentary and fragmented, with no large cavity or global blowout. Star formation plateaus, not because feedback disrupts the gas, but because the persistent perpendicular field—together with impact-seeded turbulence—prevents further mass gathering into high-density hubs, keeps the layer only marginally bound and dispersive, and hinders the development of coherent feedback bubbles. Gas, therefore, remains abundant in the midplane, while additional star formation is effectively quenched. In contrast to the 1x perpendicular case, where magnetic confinement fosters a coherent midplane slab that eventually fragments into a central cluster, triggering a blowout via stellar feedback, the 5x perpendicular run lacks such a phase transition. In 1x, feedback explosively disrupts the dense gas layer after a phase of clustered collapse; in 5x, feedback remains weak and spatially dispersed. The magnetic field's perpendicular orientation is central in both cases, but at high velocity, it combines with turbulent pressure to suppress collapse entirely, yielding a system where gas remains but stars no longer form. Thus, while the 1x run is feedback-limited, the 5x run is magnetically and kinematically quenched, highlighting the distinct regulation mechanisms introduced by collision speed.

\begin{figure*}
    \centering
    \includegraphics[width=2.0\columnwidth]{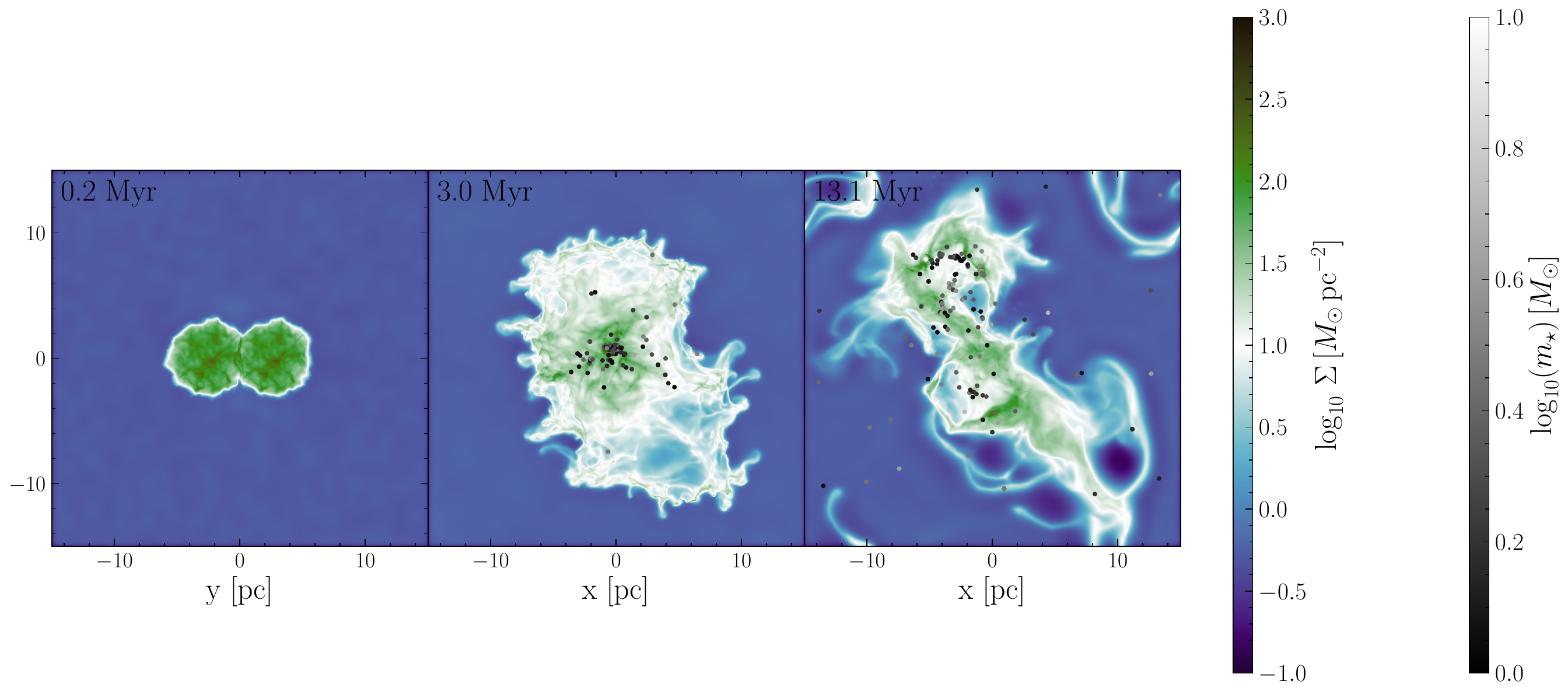}
    \caption{Projected gas surface density evolution for the 5x perpendicular magnetic field collision scenario at three representative evolutionary stages (early, intermediate, and late times). Each panel highlights how the perpendicular magnetic field shapes the morphology, initially stabilizing the gas layer against fragmentation and ultimately facilitating a dramatic feedback-driven dispersal in later stages.}
    \label{fig:5xproj}
\end{figure*}

\begin{figure*}
    \centering
    \includegraphics[width=2.0\columnwidth]{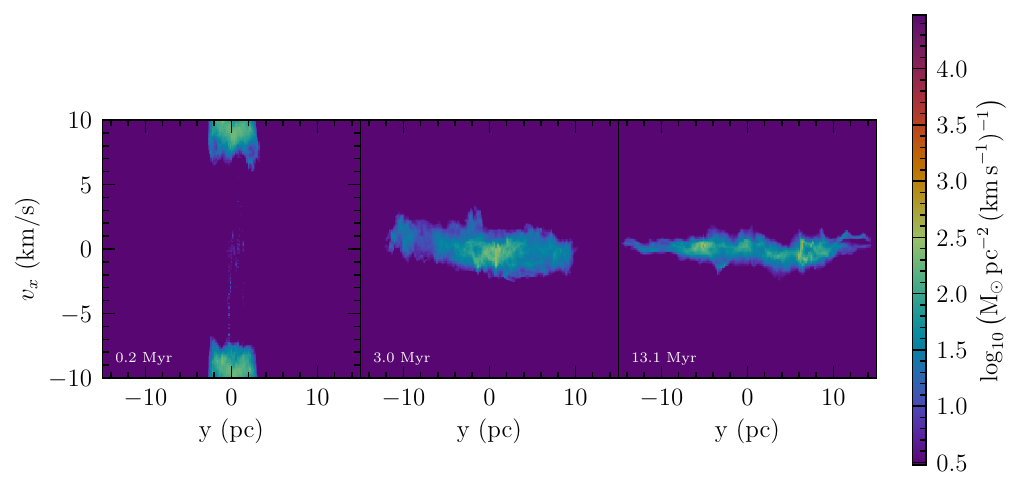}
    \caption{Position–velocity diagrams for the 5x perpendicular magnetic field collision scenario at the same evolutionary stages depicted in \autoref{fig:5xproj}. These diagrams illustrate how the velocity structure evolves from two initially distinct clouds through turbulent mixing in the intermediate stage to a highly dispersed, feedback-influenced velocity field at late times, highlighting the complex interplay between magnetic fields, gas dynamics, and stellar feedback.}
    \label{fig:5xpv}
\end{figure*}

We now examine the dynamical evolution of the 5x simulation using position–velocity (PV) diagrams, shown in \autoref{fig:5xpv}. At the early stage, the two clouds are still distinct and approaching each other; in PV this appears as two narrow vertical bands with minimal internal dispersion and little emission between them, indicating the shock interface has not yet formed. The magnetic field, initially transverse to the motion, is uncompressed and plays no significant role—clean pre-collision kinematics. By the intermediate stage, the collision has produced a turbulent shocked slab at the interface, evident as a broadened horizontal band spanning a wide velocity range, a signature of strong internal turbulence and shock-driven mixing. The perpendicular field, now compressed, enhances pressure support and channels flows laterally, suppressing collapse perpendicular to the slab; there are no clear signs of gravitational infall or feedback-driven acceleration, and chaotic mixing dominates. At the late stage, the PV diagram remains broadly similar, with a coherent, turbulent slab of moderate dispersion and no expanding shells or breakout. The velocity structure evolves little; star formation has plateaued and feedback does not manifest dynamically, while the perpendicular field continues to confine motions and suppress strong expansion or dispersal—consistent with a quasi-steady, feedback-limited regime set by the initial collision and field geometry.

\textbf{1z parallel}

\begin{figure*}
    \centering
    \includegraphics[width=2.0\columnwidth]{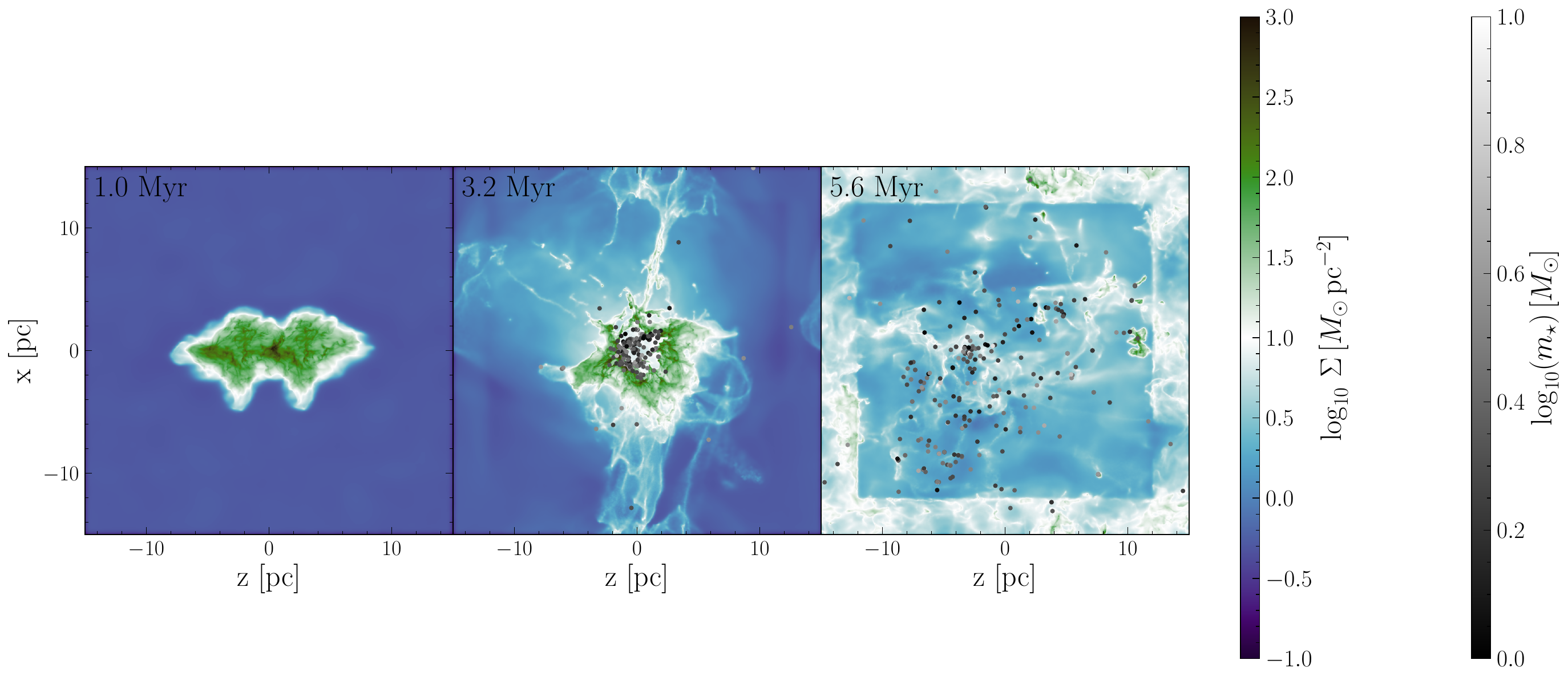}
    \caption{Projected gas surface density evolution for the 1z parallel magnetic field collision scenario at three representative evolutionary stages (early, intermediate, and late times). Each panel highlights how the perpendicular magnetic field shapes the morphology, initially stabilizing the gas layer against fragmentation and ultimately facilitating a dramatic feedback-driven dispersal in later stages.}
    \label{fig:1zproj}
\end{figure*}

In the 3\,km\,s$^{-1}$ parallel-field simulation shown in \autoref{fig:1zproj}, the clouds initially approach along the $z$-axis, forming a narrow, dense slab upon collision. The parallel magnetic field offers minimal resistance to inflow, allowing gas to stream freely and collapse rapidly along the collision axis. By 3\,Myr, a dominant midplane filament has formed, hosting a tightly clustered population of stars. Magnetic guidance channels gas toward the central slab, enhancing accretion and promoting a compact, elongated morphology. At 4.6\,Myr, feedback begins to carve out ionised cavities above and below the ridge, yet the central filament remains coherent. Compared to the perpendicular-field case, feedback-driven clearing is more gradual and aligned with the flow, reflecting the reduced magnetic opposition along the collision axis.

\begin{figure*}
    \centering
    \includegraphics[width=2.0\columnwidth]{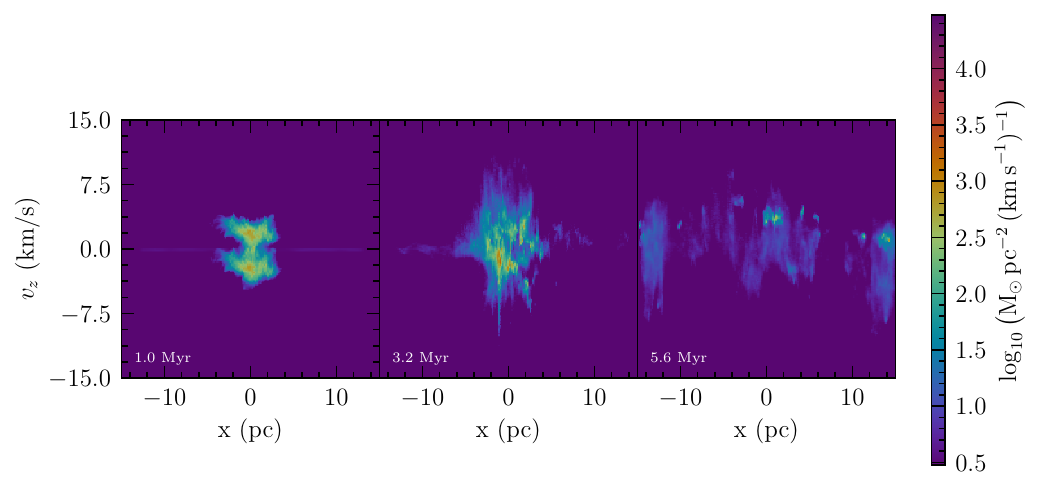}
    \caption{Position–velocity diagrams for the 1z parallel magnetic field collision scenario at the same evolutionary stages depicted in \autoref{fig:1zproj}. These diagrams illustrate how the velocity structure evolves from two initially distinct clouds through turbulent mixing in the intermediate stage to a highly dispersed, feedback-influenced velocity field at late times, highlighting the complex interplay between magnetic fields, gas dynamics, and stellar feedback.}
    \label{fig:1zpv}
\end{figure*}

Looking at the dynamical evolution of the 1z simulation using position--velocity (PV) diagrams, as shown in \autoref{fig:1zpv}. At early times (1.0 Myr), the PV structure shows a broad, symmetric feature near $x=0$ and spanning $v_z\approx\pm5\,\mathrm{km\,s^{-1}}$, corresponding to the early interaction zone where the clouds have already begun to overlap. A continuous velocity bridge is already present, indicating early-stage shock compression and coherent inflow along the parallel magnetic field. This morphology suggests rapid momentum exchange and mixing at the collision interface with minimal resistance from the field. By 3.2 Myr, the PV structure becomes broader and more vertically extended, with overlapping emission concentrated near $v_z\approx0$, but the distinct cloud components are no longer clearly separable. The velocity dispersion has increased, and the structure appears turbulent and centrally concentrated, characteristic of strong post-shock accumulation. The magnetic field continues to guide gas inflow along the collision axis, leading to a focused buildup at the midplane. At the final stage (5.6 Myr), the PV diagram fragments into streaky, disordered components spanning a wide velocity range ($|v_z|\gtrsim10\,\mathrm{km\,s^{-1}}$). These features likely trace feedback-driven expansion, residual infall, and disrupted gas flows around the forming stellar cluster. Compared to the perpendicular-field case, this evolution appears more stratified and centrally anchored, with feedback and turbulence channeled along the collision axis. The persistent central velocity structure highlights the role of the parallel field in enabling sustained inflow and delaying lateral dispersal.

\textbf{5z parallel}

\begin{figure*}
    \centering
    \includegraphics[width=2.0\columnwidth]{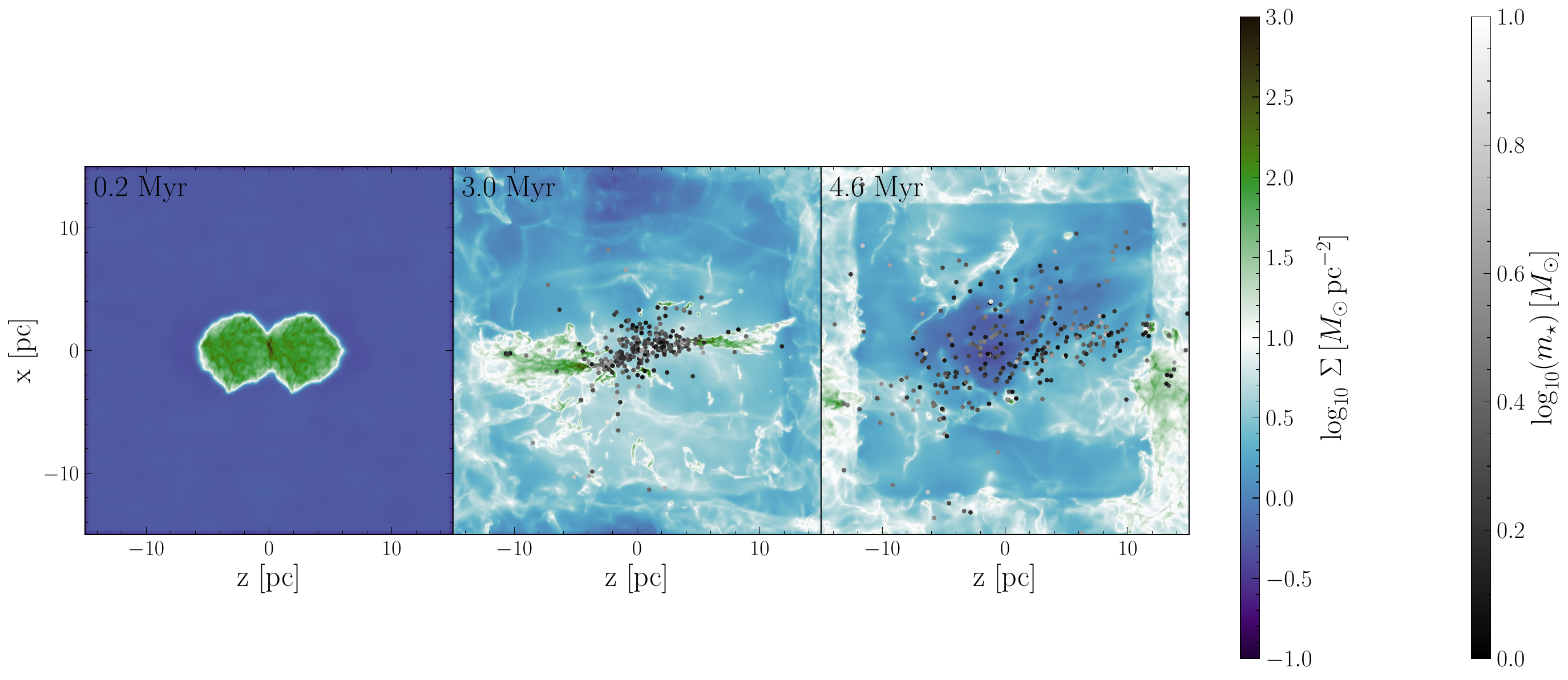}
    \caption{Projected gas surface density evolution for the 5z parallel magnetic field collision scenario at three representative evolutionary stages (early, intermediate, and late times). Each panel highlights how the perpendicular magnetic field shapes the morphology, initially stabilizing the gas layer against fragmentation and ultimately facilitating a dramatic feedback-driven dispersal in later stages.}
    \label{fig:5zproj}
\end{figure*}

In this simulation, the clouds collide at a relative speed of 15 kms$^{-1}$, with the magnetic field aligned parallel to the collisional axis. As shown in \autoref{fig:5zproj}, the clouds initially appear as two compact, nearly spherical structures converging along the $z$-axis. By 3.0 Myr, the interaction forms a dense, sheet-like midplane where stars rapidly begin to form in a concentrated ridge. Compared to the 1z case, this central slab is more fragmented, reflecting the stronger shock and faster accumulation. The morphology is dominated by filaments interspersed with clustered star particles, but remains confined to the collision axis due to the lack of magnetic tension across the slab. By 4.6 Myr, feedback from massive stars begins to disrupt the dense layer: gas is expelled outward, cavities form near the central ridge, and the stellar distribution broadens as clusters expand beyond their formation sites. While the midplane remains a focal point, the coherence of the original slab is diminished. The overall evolution reflects a rapid progression from compact ridge to dispersed structure, shaped by early compression and later feedback acting unimpeded by magnetic confinement.

\begin{figure*}
    \centering
    \includegraphics[width=2.0\columnwidth]{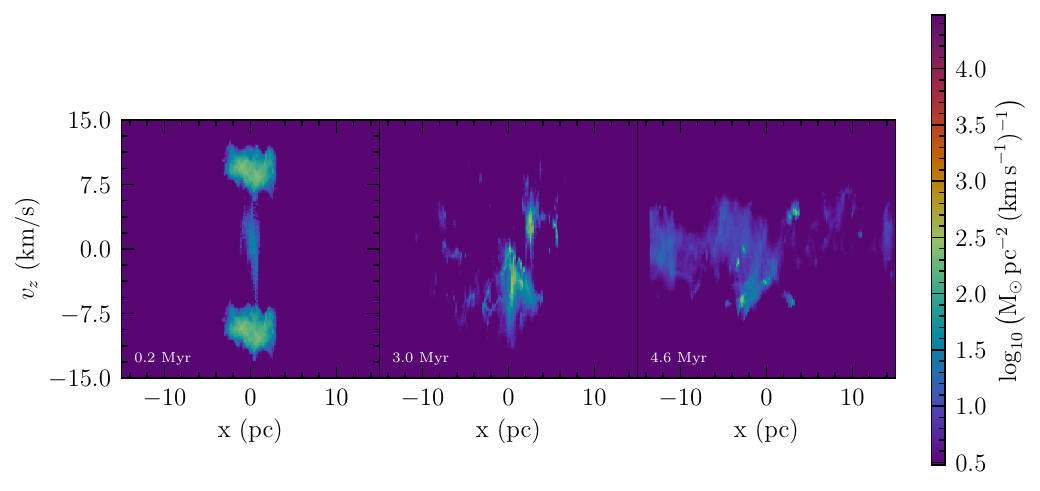}
    \caption{Position–velocity diagrams for the 5z parallel magnetic field collision scenario at the same evolutionary stages depicted in \autoref{fig:5zproj}. These diagrams illustrate how the velocity structure evolves from two initially distinct clouds through turbulent mixing in the intermediate stage to a highly dispersed, feedback-influenced velocity field at late times, highlighting the complex interplay between magnetic fields, gas dynamics, and stellar feedback.}
    \label{fig:5zpv}
\end{figure*}

The position–velocity diagram in \autoref{fig:5zpv} captures the kinematic evolution of gas motion along the collision axis ($z$) by plotting $v_z$ versus $x$ across three stages. In the early frame, the diagram features two compact, symmetric lobes at high positive and negative velocities, corresponding to the two clouds moving toward each other at 15 $\mathrm{km\,s^{-1}}$. These components are sharply defined, showing minimal velocity dispersion, and reflect the initial, ballistic approach phase prior to interaction. By the intermediate stage, the clouds have collided, and the PV structure transforms dramatically. The central region becomes filled with turbulent, shocked gas, exhibiting a broad and chaotic velocity distribution. Rather than a smooth bridge feature, the diagram shows fragmented substructures spanning a wide range of velocities, indicating rapid and violent mixing. The high collision speed, coupled with the absence of transverse magnetic tension, allows efficient fragmentation and turbulent collapse, producing a wide spread of velocities at each position. In the final stage, the PV diagram becomes even more disordered. The dense central slab seen earlier has lost coherence, and the velocity field is now populated by diffuse, low-intensity features scattered across both axes. Feedback from star formation, acting without magnetic resistance along the field-aligned flow, disperses the gas into a chaotic network of filaments and clumps. Unlike slower collisions or cases with perpendicular fields, there is no residual ordered motion or centrally confined emission structure. The kinematic evolution highlights the violently dynamic nature of the 5z case, where strong shocks and unimpeded feedback lead to rapid dispersal and suppressed star formation efficiency over time.

\subsection{Star formation rates and efficiency}
In this section, we examine how the star formation efficiency (SFE) and the number of stars formed evolve in our simulations of cloud-cloud collisions. To better understand how cloud collisions regulate fragmentation and star formation, we examine the evolution of star formation efficiency (SFE) and the number of stars formed in our simulations. Together, these two metrics characterize the outcome of gravitational collapse. While SFE quantifies the fraction of gas converted into stars, the number of stars formed ($N_\star$) indicates the degree of fragmentation within the system. We quantify the star formation efficiency using the following definition:

\begin{equation}
\text{SFE} = \frac{M_{\text{stars}}}{M_{\text{stars}} + M_{\text{gas}}},
\end{equation}

where $M_{\text{stars}}$ is the total mass in stars and $M_{\text{gas}}$ is the mass of gas remaining in the system. By applying this definition, we can compare how SFE evolves under different collision conditions, particularly variations in collision velocity and magnetic field orientation, both of which influence fragmentation and star formation outcomes.

\begin{figure}
    \centering
    \includegraphics[width=1.0\columnwidth]{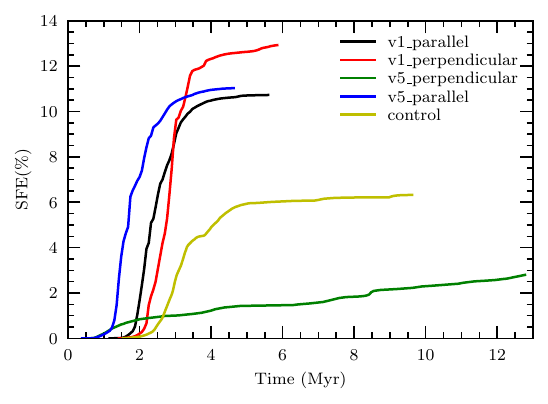}
    \caption{Evolution of star formation efficiency over time for different cloud-cloud collision scenarios.}
    \label{fig:sfe}
\end{figure}

\begin{figure}
    \centering
    \includegraphics[width=1.0\columnwidth]{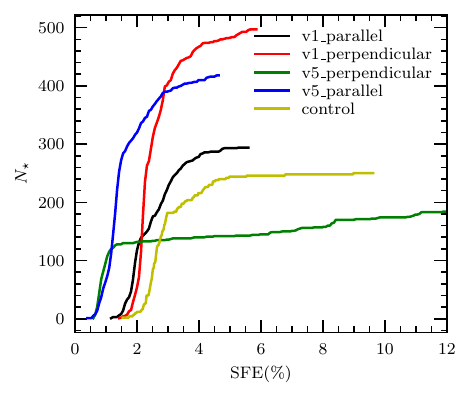}
    \caption{Number of stars formed as a function of SFE for all simulations.}
    \label{fig:nstars}
\end{figure}

\autoref{fig:sfe} and \autoref{fig:nstars} together show how the efficiency and mode of star formation depend on the combined effects of collision velocity and magnetic-field orientation. In the low-velocity collisions, the SFE rises well above the non-colliding control because the clouds spend several Myr interacting gravitationally before physical contact. This extended encounter steadily funnels gas into the interaction region, enabling persistent collapse and extensive fragmentation. The perpendicular-field case forms the largest number of stars—more than 500—as compression across field lines seeds many small-scale condensations. In contrast, the parallel-field case forms fewer (~300) but still exceeds the control, reflecting the tendency of flow-aligned fields to channel material into a smaller number of coherent, self-gravitating structures. High-velocity collisions exhibit a significantly stronger sensitivity to magnetic geometry. When the field is parallel to the flow, shock compression amplifies the magnetic pressure only weakly, allowing the dense post-shock layer to collapse rapidly. This produces an early, intense burst of star formation in which the SFE nearly doubles relative to the control, and almost 400 stars form before feedback removes the remaining dense gas and halts further fragmentation. In contrast, the perpendicular-field run represents the opposite extreme. Shock compression dramatically amplifies the transverse magnetic field, increasing magnetic pressure enough to stabilize the post-shock sheet against collapse even though the gas is strongly compressed. Both SFE and the final number of stars fall below the isolated-cloud case, with only a small, slowly forming population of stars emerging. The combination of high collision velocity and a transverse magnetic field, therefore, drives the system into a regime where magnetic support—not shock compression or self-gravity—sets the pace of collapse and limits fragmentation.

\subsection{Stellar mass distribution}

In this section, we examine the stellar mass distributions more closely, presenting stellar masses on two complementary scales: absolute and normalized by the box mass. Absolute medians highlight the characteristic mass scale of protostars under different collision conditions, while normalized medians quantify what fraction of the total cloud mass each typical star contains, directly illustrating the efficiency of mass concentration.
\begin{figure}
 \centering
 \includegraphics[width=1.0\columnwidth]{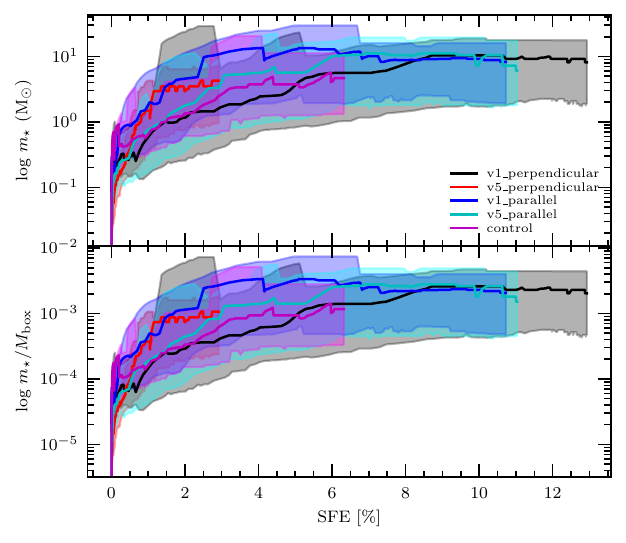}
 \caption{Evolution of mass weighted median stellar masses (solid lines) and 25th–75th percentiles (shaded bands) as a function of SFE for cloud-cloud collision runs.}
 \label{fig:median}
 \end{figure}
\autoref{fig:median} shows the evolution of median stellar masses for each collision run and the control. We characterise the stellar population using medians and percentiles computed by mass rather than by number because number-weighted statistics are highly sensitive to the abundance of the lowest-mass stars, which dominate by count but contribute little to the total mass. Small fluctuations in this population can shift number-weighted percentiles without changing the overall mass distribution. Mass-weighted measures are far more stable and reflect how the bulk of the stellar mass is distributed, making them a more robust and physically meaningful way to compare the characteristic masses across different collision setups. For all cases, median masses stabilize as simulations approach their final star formation efficiency (SFE). In the control run, star formation proceeds gradually, and the median stellar mass reflects a characteristic scale defined by unperturbed fragmentation. By comparing collisional runs to this baseline, we discern how collision speed and magnetic field orientation together influence median stellar masses. Across all runs, median stellar masses increase steadily with SFE and plateau at their final efficiencies. Runs with parallel magnetic fields consistently produce higher median masses—both in absolute and normalized terms—than their perpendicular counterparts at the same collision speed. This effect is especially pronounced at 1kms$^{-1}$, where the parallel-field run yields the highest median and widest percentile range, whereas the perpendicular-field run produces the lowest mass scale. The control simulation typically occupies an intermediate position, below parallel-field runs but generally above the high-velocity perpendicular-field run. The normalized panel confirms these differences reflect intrinsic variations in fragmentation and accretion, as all clouds begin with identical initial gas masses. Notably, the most suppressed scenario is the high-velocity perpendicular-field run, whose median mass even falls below the control. Here, the combination of rapid shock compression and strong transverse magnetic pressure limits stellar growth, favoring low-mass fragments and prematurely halting accretion. Conversely, the slower parallel collision channels material efficiently along field lines, promoting sustained growth into fewer, more massive stars. These trends reinforce that field geometry and collision dynamics jointly shape fragmentation and influence the efficiency with which mass is concentrated into individual stars, offering a clearer physical picture than absolute masses alone.
\begin{figure}
 \centering
 \includegraphics[width=1.0\columnwidth]{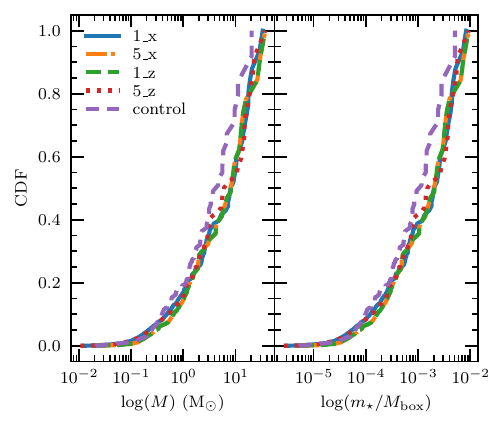}
 \caption{Cumulative distribution functions (CDFs) of mass weighted stellar mass at final SFE for collision and control runs, presented as absolute masses (left panel) and masses normalized by total cloud mass (right panel).}
 \label{fig:IMF}
 \end{figure}
The cumulative distribution functions (CDFs) in \autoref{fig:IMF} further illustrate how gas fragmentation and accretion vary under these conditions. The general similarity among distributions across all runs indicates that fundamental processes, accretion, fragmentation, and feedback, operate consistently, regardless of magnetic field orientation or collision velocity. Nonetheless, subtle yet meaningful differences emerge. Magnetic field orientation introduces weak systematic shifts: parallel-field scenarios slightly favour more massive stars, as gas preferentially flows along field lines, enhancing accretion. Conversely, perpendicular fields marginally enhance fragmentation, producing a slightly higher fraction of lower-mass stars. The control run distinctly departs from all collisional scenarios, exhibiting a systematically steeper CDF and thus producing proportionally more low-mass stars. Collision velocity, however, does not significantly influence the final mass distribution: high-velocity (5x, 5z) and low-velocity (1x, 1z) cases yield nearly identical CDF shapes. This contrasts with the pronounced influence velocity exerts on star formation efficiency and total star count, suggesting stellar masses are predominantly set by internal accretion processes rather than the collision conditions. Likewise, radiation feedback, while regulating star formation by limiting accretion and gas retention, does not substantially reshape the relative mass distribution. Overall, these results demonstrate that the stellar mass function is resilient to environmental variations compared to metrics such as star formation efficiency or the number of stars formed, with magnetic orientation causing only modest variations and collision velocity and feedback having even weaker effects.

\subsection{Feedback-driven gas dispersal and H\,II region expansion}

\begin{figure}
 \centering
 \includegraphics[width=1.0\columnwidth]{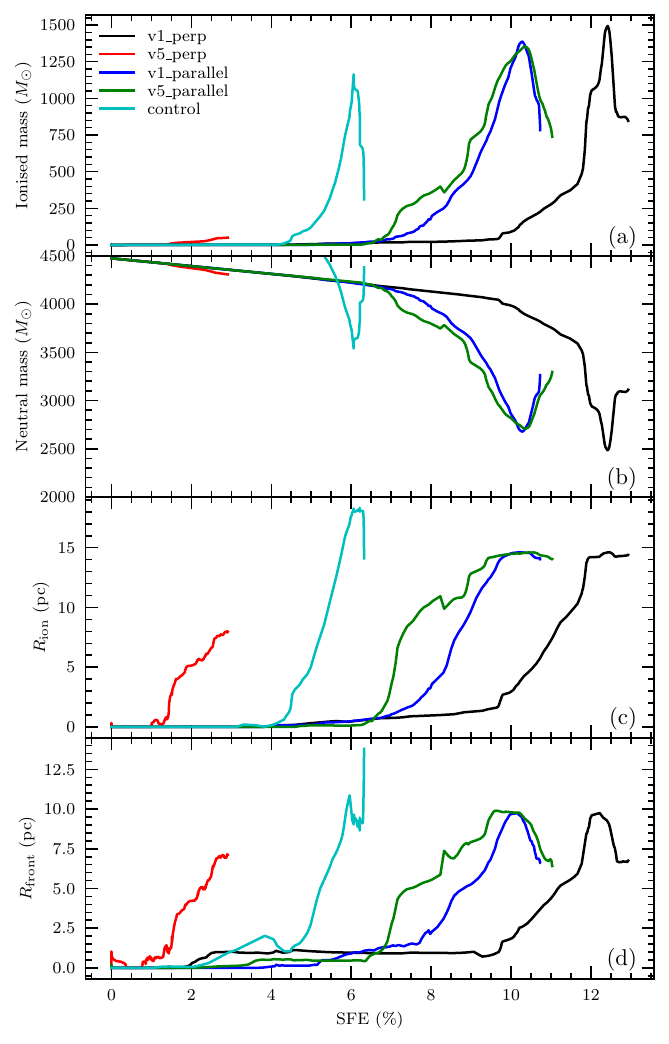}
 \caption{Evolution of key cloud gas metrics as a function of star formation efficiency (SFE) in the collision runs. Panels show (from top to bottom): (a) ionized gas mass, (b) neutral gas mass, (c) ionised region radius $R_{\rm ion}$, and (d) ionization front radius $R_{\rm front}$.}
 \label{fig:ion}
 \end{figure}

In this section, we investigate how radiation feedback reshapes the gas content and drives the expansion of the H\,II region in the collision simulations. \autoref{fig:ion} shows the evolution of four key gas metrics -- ionised gas mass, neutral gas mass, ionised region radius, and ionisation front radius -- as a function of SFE for each collision run, along with the non-colliding control case. These diagnostics trace how efficiently radiative feedback disperses gas and halts star formation, and reveal how this efficiency depends on both collision speed and magnetic geometry.

At high collision speeds, feedback activates early but behaves differently depending on field orientation. In the fast, parallel-field run, massive stars form rapidly and drive a classic ``champagne-flow'' breakout: the ionised gas mass $M_{\rm ion}$ jumps to $\sim10^3\,M_\odot$, the H\,II region expands to $R_{\rm ion}\sim12$--14\,pc, and the neutral gas reservoir collapses by $\mathrm{SFE} \approx 10\%$. This rapid dispersal closely mirrors the non-colliding control run, where, in the absence of shock compression or magnetic support, a massive star forms at $\mathrm{SFE} \approx 3$--4\%, promptly ionising $\sim(1$--$1.5)\times10^3\,M_\odot$ of gas and inflating an H\,II region to $R_{\rm ion}\sim15$--17\,pc by $\mathrm{SFE} \approx 6\%$. Both cases show that when feedback is able to act unimpeded, cloud clearing proceeds efficiently and quickly shuts down further star formation.

By contrast, in the fast, perpendicular-field run, feedback is strongly suppressed. The ionised mass remains below $\sim10^2\,M_\odot$, the ionisation front stalls at $R_{\rm front}\sim7$\,pc, and most of the cloud remains neutral and gravitationally bound. Compared to the control, this starkly illustrates how magnetic geometry can prevent feedback from coupling effectively to the cloud gas -- despite similar stellar content, the feedback is channelled into narrow escape routes and fails to clear the cloud.

At lower collision speeds, feedback still succeeds in dispersing the cloud, but over longer timescales and at higher SFE. In the slow, parallel-field run, the first massive stars appear around $\mathrm{SFE} \approx 5$--7\%, launching an H\,II region that grows steadily to $R_{\rm ion}\sim12$--13\,pc while ionising on the order of $10^3\,M_\odot$ of gas. The neutral gas mass declines gradually, reaching near-zero by $\mathrm{SFE} \approx 10$--12\%. The slow-perpendicular run shows the most extreme delay: no significant ionisation occurs until $\mathrm{SFE} \approx 8$--9\%, at which point a confined ($\sim$1--2\,pc) H\,II region suddenly bursts outward. This triggers an explosive clearing event, pushing $M_{\rm ion}$ up to $\sim1.5\times10^3\,M_\odot$ and $R_{\rm ion}$ beyond 14\,pc -- the largest of any collision case. Unlike the smooth, early dispersal seen in the control, this case underscores how a strong transverse field can trap feedback until the ionising luminosity becomes high enough to overcome magnetic confinement, resulting in a violent, late-stage blowout.


\subsection{The role of magnetic field orientation in cloud collision}

We now investigate how magnetic field orientation influences the cloud collision outcome by examining its role in shaping the post-shock energy partition and regulating the efficiency of stellar feedback. By tracking the evolution of the magnetic energy to kinetic energy ratio, magnetic energy to gravitational energy ratio, magnetic energy to internal energy ratio and magnetic energy to total energy ratio as a function of star formation efficiency (SFE), we can assess how the alignment of magnetic fields—either parallel or perpendicular to the collision axis—modulates the energy balance within the cloud, alters the feedback–gas coupling, and ultimately governs the star formation process.

\begin{figure}
 \centering
 \includegraphics[width=1.0\columnwidth]{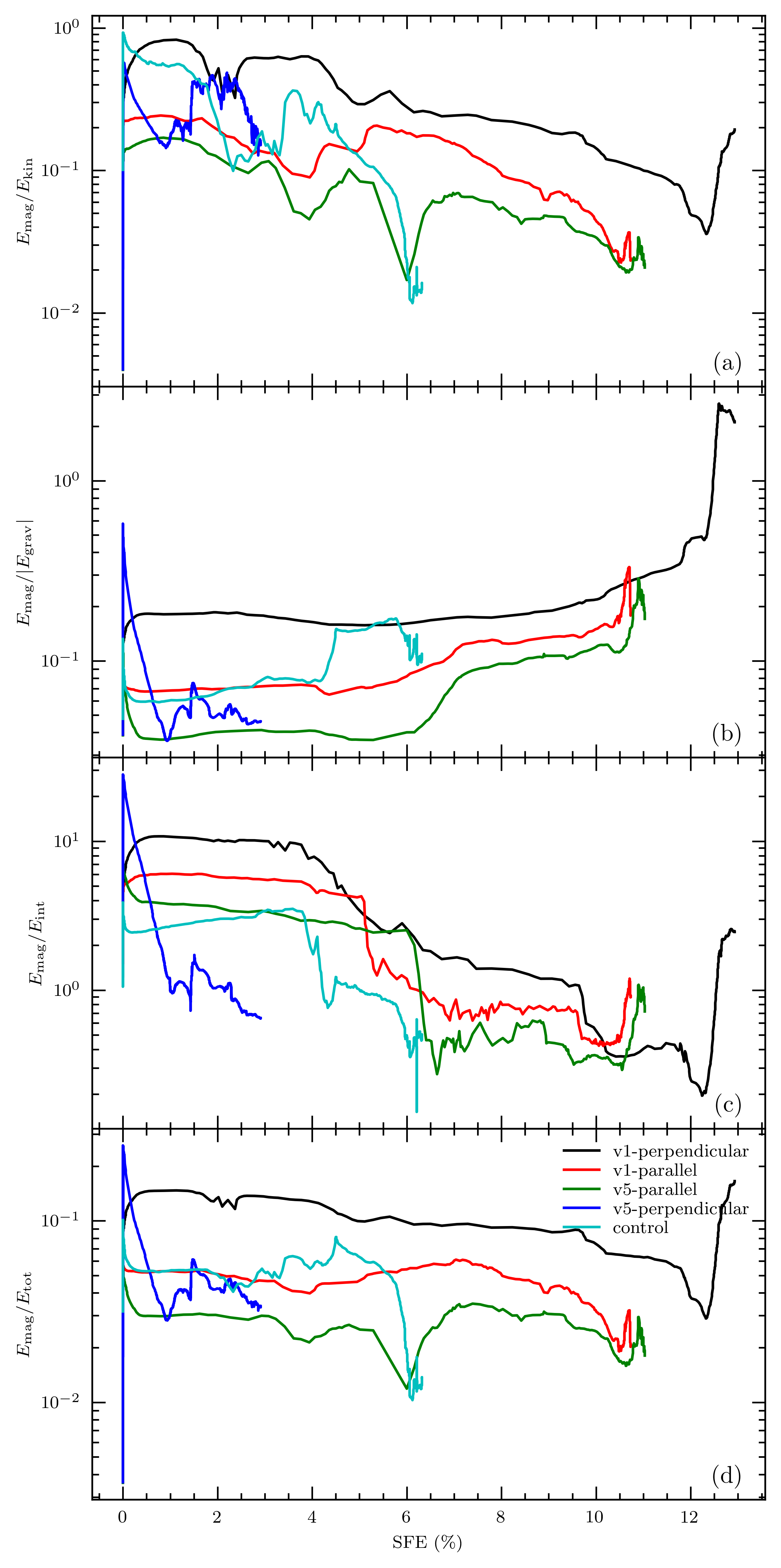}
 \caption{Evolution of various magnetic energy ratios as a function of star formation efficiency (SFE). From top to bottom, panels show magnetic-to-kinetic, magnetic-to-gravitational, magnetic-to-thermal, and magnetic-to-total energy ratios. Different curves represent the simulation runs with varying collision velocities and magnetic field orientations.}

 \label{fig:energy}
 \end{figure}

As shown in \autoref{fig:energy}, the tracks separate in a highly ordered way across the four panels. In panel (a), the v1-perpendicular (black), v1-parallel (red), and control (cyan) runs define the upper envelope with \(E_{\rm mag}/E_{\rm kin} \sim 10^{-1}\), indicating that turbulent kinetic energy is always comparable to, or weaker than, the magnetic reservoir, whereas the v5-parallel case (green) sits at the lower envelope with \(E_{\rm mag}/E_{\rm kin} \sim\) a few \(\times 10^{-2}\), reflecting a strong shock that maintains kinetic dominance and only modest field amplification; the v5-perpendicular run (blue) lies in between, with \(E_{\rm mag}/E_{\rm kin}\) settling around a few \(\times 10^{-2}\)–\(10^{-1}\) once star formation begins. Panel (b) shows the corresponding ratio to gravity: the v1 runs and the control decline gently from \(E_{\rm mag}/|E_{\rm grav}|\sim 0.1\), consistent with a gradual deepening of the potential well, whereas v5-parallel drops rapidly to \(\sim 10^{-2}\), signalling that self-gravity quickly overwhelms the magnetic field as gas collapses efficiently along the field lines; by contrast, v5-perpendicular remains several times higher than v5-parallel at a given SFE, demonstrating that the shock-compressed transverse field keeps the slab only marginally more self-gravitating than magnetically supported. The thermal balance in panel (c) follows the same hierarchy. In the v1 runs and the control, \(E_{\rm mag}/E_{\rm int}\) stays of order unity or larger, indicating mild heating and efficient cooling; v5-parallel again sits lowest with \(E_{\rm mag}/E_{\rm int} \sim 0.3\)–0.5, because the high-Mach collision thermalises much of the inflow; v5-perpendicular stabilises near \(E_{\rm mag}/E_{\rm int}\sim 1\), so magnetic and thermal pressures remain comparable. Finally, panel (d) shows that the v1 runs and the control retain the highest magnetic fractions, \(E_{\rm mag}/E_{\rm tot}\sim 0.05\)–0.1, whereas v5-parallel is always lowest (\(\lesssim 0.02\)), with v5-perpendicular again at an intermediate, nearly flat \(\sim 0.03\). Crucially, it is this intermediate regime—moderate \(E_{\rm mag}/E_{\rm kin}\), \(E_{\rm mag}/E_{\rm int}\approx 1\), and a slowly declining \(E_{\rm mag}/|E_{\rm grav}|\) at a roughly constant \(E_{\rm mag}/E_{\rm tot}\)—that corresponds uniquely to the v5-perpendicular run, which ends at a final SFE of only \(\sim 2\) per cent. In that configuration the perpendicular fast shock amplifies the field enough to prevent a v5-parallel-like runaway collapse (low magnetic ratios, gravity-dominated), but the resulting magnetically stiff, moderately warm slab also prevents the v1-like outcome where feedback can eventually blow out the gas at SFE \(\sim 10\)–12 per cent: outflows are confined to narrow chimneys, most of the gas remains neither deeply bound nor efficiently cleared, and star formation simply stalls in a magnetically regulated, weakly star-forming state.

\section{Discussion}

Our results highlight how cloud–cloud collisions significantly influence star formation, particularly the formation of massive stars, with outcomes dependent on collision dynamics, magnetic fields, and stellar feedback. Head-on cloud collisions accelerate star formation compared to isolated clouds, consistent with observational indications. For instance, \cite{2020MNRAS.494..246T} demonstrated that collisions between gravitationally bound clouds roughly double the stellar mass formed compared to non-colliding clouds. Our low-velocity collisions, where clouds remain bound through prolonged interaction, achieved star formation efficiencies approximately twice as high as control runs, supporting the notion that converging flows promote dense core formation beyond turbulent fragmentation alone.

Collision velocity is one of the key factors in determining star formation outcomes. In slower collisions, the prolonged interaction allows mutual gravity to accumulate gas centrally before any disruptive feedback can set in. This results in sustained star formation with high efficiency. These low-velocity runs produce prominent filamentary structures and hub-like gas concentrations along the collision interface, consistent with earlier results from bound–bound collisions  \citep{2020MNRAS.494..246T}.  Such hub-filament systems are typically associated with massive star formation regions, where converging filaments feed a central proto-cluster. Our simulations indeed reveal dense filament networks funneling material into central hubs, consistent with this picture. Compared ot slow collisions, high-velocity collisions lead to rapid and violent shock compression that quickly raises gas densities. However, the shock also generates strong turbulence and thermal pressure in the post-shock layer. Without magnetic support, such collisions can trigger an intense but short-lived starburst. In our high-speed parallel-field case, the star formation efficiency surged to nearly double that of the control run within just 2–3 Myr. Yet the resulting feedback from massive stars rapidly disrupted the shocked gas, causing star formation to plateau. This outcome suggests that there may be an optimal collision speed: if too slow, the interaction resembles typical turbulent fragmentation; if too fast, the induced starburst is quenched before sustained cluster formation. \cite{10.1093/mnras/stab150} found that the formation of massive clusters likely requires both high gas density and collision speeds exceeding $\ approx$20 $\mathrm{km\,s^{-1}}$, implying that only the most energetic collisions in dense environments yield massive clusters. Our 15 $\mathrm{km\,s^{-1}}$ collisions—moderate by observed standards—did produce more stellar mass than the no-collision case, but the gains were ultimately constrained by rapid gas expulsion. Collision velocity also influences the spatial morphology of the post-collision cloud. High-speed impacts tend to either disperse clouds entirely or concentrate them into compact remnants. In contrast, slower collisions build up more extended, filament-rich structures. This trend is consistent with \cite{2023MNRAS.522..891W}, who report that increasing the collision velocity reduces the spatial extent of the remnant, leading to fewer but more massive clusters.

Collision velocity also influences the spatial morphology of the post-collision cloud. High-speed impacts tend to either disperse clouds entirely or concentrate them into compact remnants, whereas slower collisions build up more extended, filament-rich structures. Observationally, cloud-cloud collisions often produce characteristic kinematic signatures known as velocity bridges, first identified clearly by \citet{2015MNRAS.450...10H}. These velocity bridges manifest as intermediate-velocity gas connecting two distinct velocity components and have since been observed in numerous massive star-forming regions, including Westerlund 2 \citep{2009ApJ...696L.115F}, NGC 3603 \citep{2014ApJ...780...36F}, and RCW 120 \citep{2015ApJ...806....7T}. Our simulations robustly reproduce such features, suggesting that velocity bridges observed in these regions can indeed be indicative of cloud collisions. Future high-resolution observational campaigns targeting these intermediate-velocity features can therefore directly test and validate our simulated collision scenarios.

Magnetic field orientation also plays an important role in the outcome of cloud collisions. In the parallel-field configuration, the field offers minimal resistance to converging gas flows, allowing gravity and ram pressure to dominate. The post-shock region becomes dominated by gravitational collapse, forming a dense, flattened slab or filament. In our 3 $\mathrm{km\,s^{-1}}$ parallel-field run, a single massive filament formed along the collision axis, giving rise to a tight cluster of sink particles. This scenario leads to coherent collapse and the formation of fewer but more massive fragments, reflected in a higher median stellar mass. A perpendicular magnetic field configuration shows a different evolution. The collision front compresses the magnetic field lines, amplifying magnetic pressure in the shock interface. This pressure stabilizes the post-shock layer against gravitational collapse, at least temporarily. Perpendicular-field collisions tend to form broader, magnetically supported slabs characterized by intricate networks of filaments and clumps. Star formation in these runs is more distributed and delayed. In our 3 $\mathrm{km\,s^{-1}}$ perpendicular-field run, hundreds of low-mass stars formed along an extended ridge of dense gas before any dominant cluster emerged. Fragmentation was channeled along field-aligned filaments, resulting in a lower characteristic stellar mass. These behaviors align with the findings of \cite{2013ApJ...774L..31I}, who showed that perpendicular fields promote filamentary structure while delaying collapse. \cite{2023MNRAS.522.4972S} also found that magnetic fields in high-speed collisions have a dual role: they initially help mass accumulate in the shock layer but later hinder collapse by expanding and dispersing the compressed gas. In our high-velocity perpendicular run, this magnetic effect was extreme—shock-compressed magnetic pressure prevented the formation of any bound cluster. The shocked layer expanded laterally and remained turbulent, with star formation almost completely suppressed. This outcome was not observed in non-magnetized runs, emphasizing the suppressive role of strong perpendicular fields. Sakre et al. noted that magnetic support in such collisions can destroy nascent cores or halt their growth, especially if the column density is too low to counterbalance the magnetic pressure. Our results support this view: the 15 $\mathrm{km\,s^{-1}}$ perpendicular-field case resulted in the immediate dispersal of the shocked gas. At the same time, the parallel-field equivalent did form a massive cluster before feedback terminated the process.

Radiative feedback in our simulations acts as a self-regulating mechanism, capping the star formation efficiency at ~10–15\%. This range aligns with both observed star formation efficiencies in molecular clouds and results from radiation-hydrodynamic simulations by \cite{2022ApJ...935...53H}, who found that radiative feedback roughly halved the final stellar mass compared to no-feedback scenarios. In our low-velocity perpendicular-field runs, magnetic confinement allowed a large gas reservoir to accumulate before feedback initiated a dramatic blow-out phase. Once the central massive stars formed, their ionizing radiation carved out a spherical HII region that burst through the magnetically supported layer, expelling much of the remaining gas. This sequence resembles a “trigger then blow-out” model, akin to what is thought to occur in super bubbles or giant HII regions such as 30 Doradus. In contrast, the high-speed parallel collision saw more immediate feedback effects: with less magnetic confinement, ionized gas escaped early along the collision axis, leading to a slower, asymmetric dispersal. These scenarios suggest that magnetic field orientation modulates how feedback energy couples to the cloud. Perpendicular fields tend to bottle up energy until catastrophic breakout occurs, while parallel fields allow for more continuous gas venting. Regardless of geometry, radiative feedback terminated star formation well before gas exhaustion, reinforcing the need to include feedback in any realistic model. Earlier hydrodynamic studies without feedback (e.g. \cite{2017ApJ...835..137W}) reported unrealistically high star formation rates due to unchecked gas collapse [10]. Feedback-driven structures such as expanding shells, cavities, and ionization fronts—often observed in regions like RCW 120—are best explained when feedback is both included and spatially modulated by magnetic conditions.

Despite significant variation in global outcomes (e.g., total stellar mass, cluster size), the initial mass function (IMF) in our simulations remains broadly consistent across runs. Parallel-field collisions yielded slightly higher median stellar masses due to coherent collapse into massive filaments, whereas perpendicular-field collisions—especially at low speed—produced a larger number of low-mass stars owing to widespread, magnetically channeled fragmentation. These differences are relatively modest, and cumulative mass distributions remain largely congruent across cases. This robustness is encouraging, as it suggests that even when star formation is triggered by cloud collisions, the IMF is not dramatically skewed. Our results support the notion that while collisions alter where and how fast stars form, the mass of each star still emerges from localized processes such as hierarchical fragmentation and competitive accretion. Observations of massive clusters like Westerlund 2 or NGC 3603 also show IMF slopes consistent with canonical values, further supporting the idea that collision-induced clusters are not top-heavy in their mass distributions.

\section{Conclusions}

In this study, we have systematically investigated the impact of cloud–cloud collisions on star formation, focusing particularly on how varying collision velocities and magnetic field orientations influence the morphology, star formation efficiency (SFE), fragmentation, stellar mass distribution, and feedback-driven gas dispersal. Our simulations, utilizing the STARFORGE framework with full ideal magnetohydrodynamics and radiative feedback, reveal that cloud collisions significantly enhance star formation efficiency compared to isolated cloud scenarios, especially at lower collision velocities where prolonged gravitational interactions effectively accumulate gas for sustained star formation. The key findings from our studies are

\begin{itemize}
    \item Collision speed crucially affects both the star formation efficiency and cloud morphology. Lower-velocity collisions facilitate the development of coherent, elongated filaments and hub-like structures that lead to sustained, high-efficiency star formation. In contrast, high-velocity impacts rapidly compress gas but also produce strong turbulent motions and feedback-driven dispersal, which can prematurely quench star formation after an initial burst.

    \item Magnetic field orientation profoundly influences the collision outcomes. Parallel fields allow gas to efficiently flow along magnetic lines, resulting in early gravitational collapse and the formation of fewer but more massive stellar fragments. Conversely, perpendicular fields induce magnetic pressure that initially stabilizes the shock-compressed gas against gravitational collapse, resulting in delayed, distributed star formation and lower median stellar masses.

    \item Radiative feedback from massive stars plays a pivotal regulatory role in star formation. Regardless of magnetic geometry, feedback consistently halts star formation at moderate efficiencies (~10–15\%), preventing runaway gravitational collapse. The feedback-driven dispersal processes observed, including formation of expanding H II regions and cleared cavities, strongly depend on magnetic field geometry—perpendicular fields temporarily confine feedback energy, causing dramatic late-time gas dispersal, whereas parallel fields permit more continuous and directional clearing.
\end{itemize}

Despite marked variations in global properties like total stellar mass and cloud morphology, the stellar mass function remains relatively robust across different collision scenarios. Magnetic orientation introduces modest systematic variations in mass distributions, favoring slightly more massive stars in parallel-field configurations, but collision velocity does not significantly alter the intrinsic shape of the mass function.

Our findings carry significant implications for interpreting observations of massive star-forming regions, suggesting that cloud collisions are viable mechanisms for rapidly triggering star formation. The distinct morphological and kinematic signatures found in our simulations, including filamentary structures, expanding H II regions, and turbulent velocity distributions, provide testable predictions for observational studies. Future observational efforts could aim to identify these kinematic signatures—such as velocity bridges and filamentary gas structures—across a broader sample of massive star-forming regions to further substantiate the role of cloud collisions.

Theoretically, extending simulations to incorporate additional physics such as non-ideal MHD effects (e.g., ambipolar diffusion and Hall effects), varying cloud mass ratios, and different initial turbulence patterns will further deepen our understanding of collision-driven star formation. Furthermore, exploring feedback from additional stellar processes (such as protostellar jets, stellar winds, and supernova explosions) can offer a more comprehensive picture of the lifecycle of molecular clouds in collision-rich environments. Such combined theoretical and observational advancements will refine our understanding of the complex interplay governing massive star and cluster formation in galaxies.

\section*{Acknowledgements}

Acknowledging Blizzard of Oz for helping me power through this paper. 

\section*{Data Availability}
The data underlying this article will be shared on reasonable request to the corresponding author.






\bibliographystyle{apsrev4-1}

\bibliography{oja_template}




\end{document}